\documentclass[%
floatfix,
 aip,
 amsmath,assume,
 reprint,%
]{revtex4-1}

\usepackage{graphicx}
\usepackage{dcolumn}
\usepackage{bm}

\usepackage[utf8]{inputenc}
\usepackage[T1]{fontenc}
\usepackage{mathptmx}
\usepackage{etoolbox}

\makeatletter
\def\@email#1#2{%
 \endgroup
 \patchcmd{\titleblock@produce}
  {\frontmatter@RRAPformat}
  {\frontmatter@RRAPformat{\produce@RRAP{*#1\href{mailto:#2}{#2}}}\frontmatter@RRAPformat}
  {}{}
}%
\makeatother
\begin{document}

\preprint{AIP/123-QED}

\title{{Sinusoidal Transmission Grating Spectrometer for EUV Measure}}
\author{N. Kliss}
 \email{noa.kliss@mail.huji.ac.il}
\affiliation{ 
L2X Labs, Jerusalem, Israel%
}%
 
\affiliation{The Hebrew University in Jerusalem, Israel}

\author{J. Wengrowicz}%
\affiliation{ 
L2X Labs, Jerusalem, Israel
 }%

\author{J. Papeer}
\affiliation{ 
L2X Labs, Jerusalem, Israel%
 }%

\author{E. Porat}
 \affiliation{%
KLA, Yavne, Israel
}%

\author{A. Zigler}
 \affiliation{%
The Hebrew University in Jerusalem, Israel
}%

\author{Y. Frank}
\affiliation{ 
L2X Labs, Jerusalem, Israel%
 }%

\date{\today}

\begin{abstract}
Spectral measurements play a vital role in understanding laser-plasma interactions.
The ability to accurately measure the spectrum of radiation sources is crucial for unraveling the underlying physics.
In this article, we introduce a novel approach that significantly enhances the efficiency of binary Sinusoidal Transmission Grating Spectrometers (STGS). The grating was tailored especially for Extreme Ultraviolet (EUV) measurements. The new design, High Contrast Sinusoidal Transmission Grating (HCSTG), not only suppresses high diffraction orders and retains the advantageous properties of previous designs but also exhibits a fourfold improvement in first-order efficiency. In addition, the HCSTG offers exceptional purity in the first order due to effectively eliminating half-order contributions from the diffraction pattern.
The HCSTG spectrometer was employed to measure the emission of laser-produced Sn plasma in the 1-50 nm spectral range, achieving spectral resolution of  $\lambda/\Delta\lambda=60$.
 We provide a comprehensive analysis comparing the diffraction patterns of different STGs, highlighting the advantages offered by the HCSTG design. This novel, enhanced efficiency HCSTG spectrometer, opens new possibilities for accurate and sensitive EUV spectral measurements.

\end{abstract}

\maketitle

\section{\label{sec:1}Introduction}

In recent years, there has been a notable upsurge in the utilization of Extreme Ultraviolet (EUV) radiation for lithography processes within the semiconductor chip industry. As a result, there is a growing demand in basic research in the field of EUV radiation production. Numerous studies investigating the creation processes of EUV were published in recent years e.g.\cite{Schupp2019,Banine2011,Torretti2020,Versolato2019}. 
Accurate detection and characterization of EUV light emitted by various sources, including plasma, synchrotron, and free-electron lasers, holds paramount significance for both industrial and academic purposes. Particularly, advancements in unique optical elements, such as multi-layer mirrors \cite{Voronov2011,Voronov2012}, have amplified the importance of detecting and measuring EUV radiation, especially within the critical wavelength range of around 13.5 nm. 
However, measuring EUV radiation poses challenges due to its high absorption in nearly all materials. Consequently, transmission gratings have become a prevalent choice for EUV and soft-x-ray spectroscopy. The conventional design of bar transmission gratings, consisting of parallel grooves with a square-wave transmission function \cite{Hambach2001,Golub2015,Miao2014,Hurvitz2012}, is known to encounter issues with overlapping high dispersion orders, hampering spectral measurement accuracy and limiting the width of accurately measurable spectra.
To overcome the high-order overlap effects, several designs for optical elements with 
sinusoidal transmission function were suggested including Quasi-sinusoidal TG \cite{Kuang2011,Fan2015}, zig-zag TG\cite{Zang2012} and Sinusoidal TG known as the STG \cite{Shpilman2019, Lightman2019}. 
A binary sinusoidal transmission grating is a two-dimensional periodic mask with alternating transparent and opaque regions that offer an amplitude transmission function producing only the 0, 1, and -1 orders. As presented in previous works\cite{Lightman2019, Shpilman2019}, its utilization enables the mitigation of high-order overlap issues, as the far-field dispersion contains only the first orders.
In this article, we introduce a novel design of the High Contrast Sinusoidal Transmission Grating (HCSTG) and present a comprehensive comparison with a regular STG design \cite{Lightman2019}. 
Summing over a full STG vertically reveals a sinusoidal function of the open area fraction compared to all the area, opaque and open, along the grating.
The amplitude of this sinusoidal function can be measured between 0 and 1 and will be called the "contrast" of the grating, as it determines the contrast of the sine function produced by summing over all the grating vertically. The HCSTG exhibits an impressive contrast of almost 1, a substantial improvement over previous designs, which reached only about 0.5. The increased contrast in the new design leads to a fourfold improvement in efficiency for the first diffraction order. This enhanced contrast of the HCSTG holds great promise for improving the accuracy of spectral measurements in this critical domain. This study included the development of a spectrometer based on the HCSTG design. Specifically tailored for high-resolution EUV measurements, the spectrometer's outcomes are detailed in this article.

\section{\label{sec:2}System description}

\subsection{\label{sec:2A}TG desing}

\begin{figure}
\includegraphics[width=80mm]{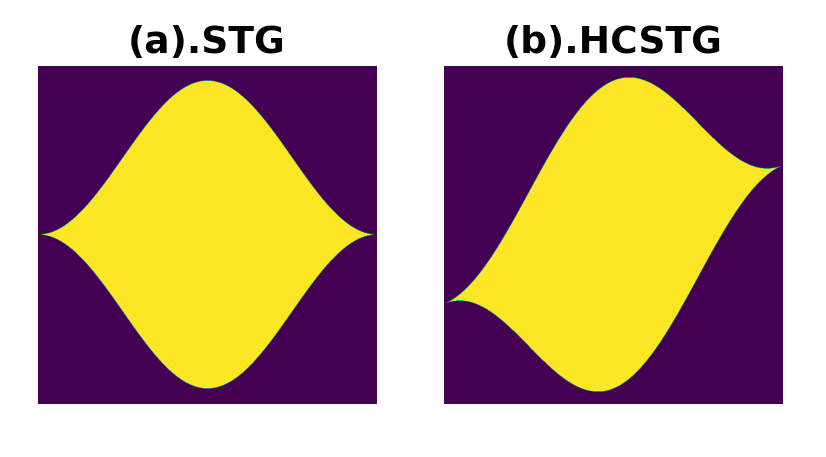}
\caption{\label{fig: one_eye} Two aperture designs. (a) a regular STG eye shape hole (b) HCSTG hole design contains a linear term that distorts the eye shape.}
\end{figure}

The STG hole outline function is:

\begin{equation}\label{1}
|y|=cos^2(x)
\end{equation}

In order to make an optical element that maintains the transfer function while enabling a reduced horizontal separation between the holes without inducing mesh fractures, an additional term was incorporated:

\begin{equation}\label{2}
|y+f(x)|=cos^2(x)
\end{equation}

In our case, for the HCSTG design, we used the simplest solution, a linear function. $f(x)=s\frac{x}{\pi}$ 
So the outline function is:

\begin{equation}[\label{3}
    \left|y+s\frac{x}{\pi}\right|=cos^2(x)
\end{equation}

When s is a parameter between zero and one. The additional linear term created a deviation of the edges of the eyes as shown in FIG.\ref{fig: one_eye}. This deviation allowed the production of a much denser grating.

\subsection{\label{sec:2B} Spectrometer Parameters} 

The HCSTG, which was described in the previous section, was utilized in a spectrometer for EUV measurement as presented in FIG.\ref{fig:system}. Further, this is explained in the experiment setup part.
The distance from the source to the target is L, which satisfies 
$$L=L_1+L_2$$
when $L_1$ is the source-to-grating distance and $L_2$ is the grating-to-sensor distance. The total distance-L was determined mostly by the radiation intensity. The radiation intensity decays approximately linearly in $L^2$. From the evaluation of the radiation emitted by the source, the upper limit for the distance L was determined in order to maintain a good signal-to-noise ratio.   
Radiation going through the transmission grating will diffract according to Bragg's low: 
\begin{equation}\label{4}
\sin(\theta)=m\frac{\lambda}{d}
\end{equation}
where $\lambda$ is the wavelength, $d$ is the period of
the grating. and $m$ is the diffraction's order. Since our grating is an STG, only the zeroth and first orders will appear hence $m=0,\pm1$. The spectral broadening can be calculated as shown in previous work\cite{Sailaja1998}, and the resolution is given by:   

\begin{equation}\label{5}
\Delta\lambda=\frac{d}{m}\sqrt{\left(\frac{\Delta s+w}{L_1}+\frac{w}{L_2}\right)^2+\left(\frac{\lambda}{w}\right)^2}
\end{equation}

where $\Delta s$ is the source size and $w$ is the grating width. From equation \ref{2}, one can see that for infinitesimal source and infinite distance the maximum resolution can be achieved and it is limited by the number of grating periods N:

\begin{equation}\label{6}
\frac{\lambda}{\Delta\lambda}\leq \frac{w}{d}=N
\end{equation}

A point source would cover height $\Delta H$ on the detector:
\begin{equation}\label{7}
\Delta H=\sqrt{\left(H\cdot\frac{L}{L_1}\right)^2+\left(\frac{\lambda\cdot L_2}{H}\right)^2}
\end{equation}

Reducing the value of $\Delta H$ leads to decreased noise during vertical spectrum integration. The smallest achievable $\Delta H$ occurs when the grating height H is:

\begin{equation}\label{8}
H=\sqrt{\frac{L_{1}\cdot L_{2}\cdot\lambda}{L}}
\end{equation}

The grating and spectrometer parameters were determined for maximal spectral resolution in the EUV regime, specifically at 13.5 nm.  

\subsection{\label{sec:2C}Experiment Setup}

\begin{figure}
\includegraphics[width=85mm,scale=0.5]{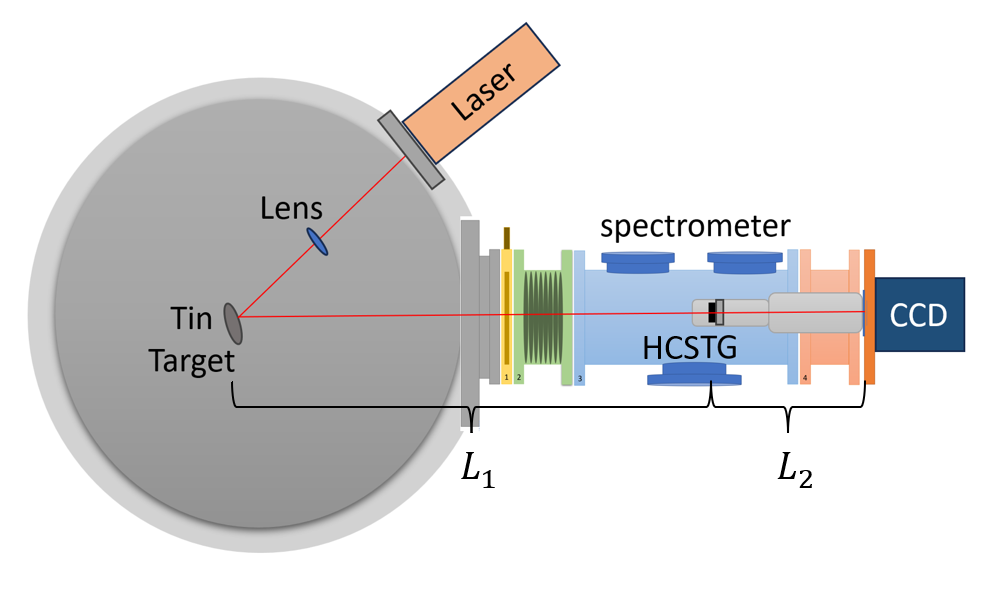}
\caption{\label{fig:system} Schematic diagram of the optical setup. }
\end{figure}

In the experiment, a Surelight ND:YAG laser operating at $\lambda=1064 nm$ with FWHM of 10 ns pulse was used to produce the plasma in a Laser Produced Plasma (LPP) interaction process. A schematic diagram of the experiment can be seen in FIG. \ref{fig:system}. The laser was focused on the Sn target
with a 15 cm focal length lens to a spot size of $\sim100 \mu m$ diameter. The EUV radiation emitted in the LPP process was measured using the HCSTG spectrometer. The spectrometer was inclined at an angle of 5 degrees relative to the normal, with respect to the surface of the target. A CCD camera was used as a sensor to detect the time-integrated emitted spectrum. All the experiments took place in a vacuum chamber with $\sim10^-6  Torr$.
The model of the camera used for the detector in the spectrometer is a Newton Andor CCD camera with a back-illuminated sensor. The CCD camera sensor is sensitive to EUV radiation, particularly around 13.5 nm. The spectral resolution of the device in wavelength of 13.5 nm is $\frac{\lambda}{\Delta\lambda}=60$.

\subsection{\label{sec:2D}Calibration Method}

\begin{figure}
\includegraphics[width=85mm]{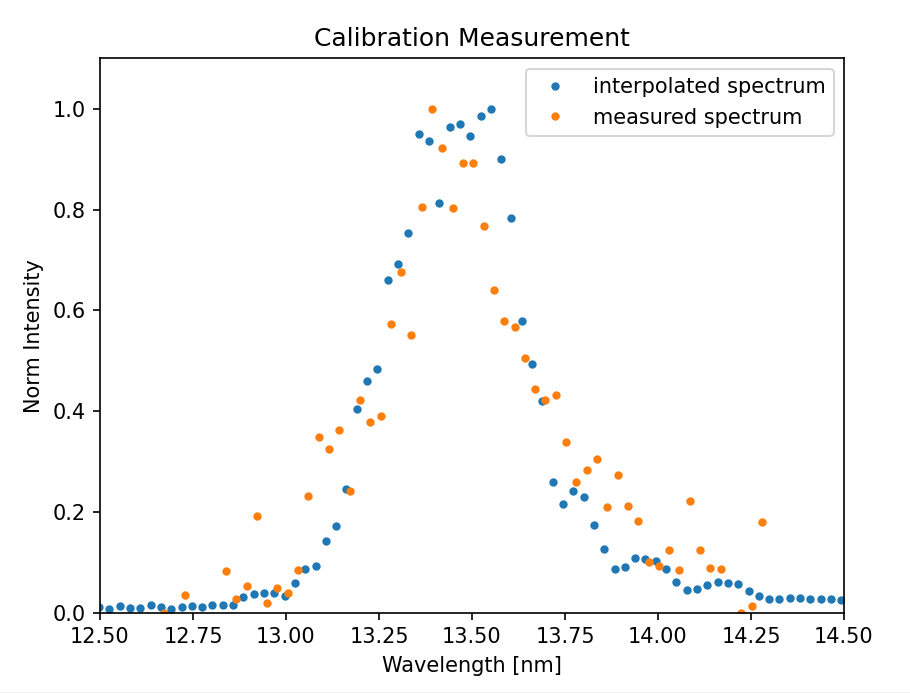}
\caption{\label{fig: calib} Measurement of a semi-monochromatic source and an interpolation of the entire spectrum with the reflection function of the multi-layer mirror}
\end{figure}

The spectrometer and the HCSTG were experimentally calibrated to confirm that the parameters $L_1, L_2, d$ are according to the design specifications. A quasi-monochromatic source was used for the calibration. The source was obtained by creating laser-produced Sn plasma and reflecting it with $5^{\circ}$ multi-layer mirror\cite{Voronov2011, Voronov2012}. The multi-layer mirror is unique because of its narrow reflection curve centered around 13.5 nm. The radiation emitted from the Sn plasma passed through the mirror into the spectrometer. The transition in the mirror cut the spectrum according to the mirror's narrow reflection curve, allowing us to see the diffraction pattern in a specific wavelength of 13.5 nm. The outcome of the measurement can be seen in FIG.\ref{fig: calib} alongside the corresponding calculated spectrum. The calculated spectrum is a composite of the entire emitted spectrum from the laser-produced Sn plasma, interpolated with the known reflecting spectrum of the multi-layer mirror. The evident congruence between the measurement and the interpolation validates the spectrometer and TG parameters.

\section{\label{sec:3}Fabrication Process}

\begin{figure}
\includegraphics[width=85mm]{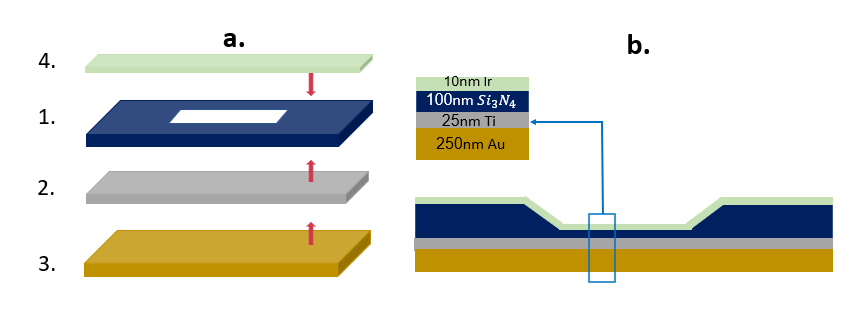}
\caption{\label{fig: fab} The STG fabrication process a. The different evaporation stages.
(a.1) The base is a TEM support film, made out of a thick silicon frame and a 100X100 micron square aperture of 100 nm $Si_3N_4$ membrane (a.2) deposition of 25 nm Ti (a.3)
deposition of 250 nm Au (a.4) deposition of 10 nm Ir b.cross section view. We can see the 
different layers. This model is put into the FIB. The FIB is used for drilling the pattern 
through all the layers to produce the final STG}
\end{figure}

The fabrication process of the sinusoidal transmission grating involves several steps. The production includes substrate preparation, deposition, drilling pattern using Focused Ion Beam (FIB), and quality assurance by Scanning Electron Microscope (SEM) 
imaging. Each of these steps must be carried out with a high degree of precision and accuracy
to ensure the quality and performance of the final product.
The basic substrate on which the deposition took place is a commercial Transmission Electron Microscopy (TEM) support film (FIG \ref{fig: fab}.a.1) . The TEM support film is a 100 nm
layer of silicon nitride ($Si_3N_4$) with a frame of silicon. The aperture is 0.1 mm X 0.1 mm square. A thin layer of 25 nm Ti is 
evaporated over the substrate (FIG \ref{fig: fab}.a.2), and another layer of 250 nm Au is 
evaporated over it (FIG \ref{fig: fab}.a.3). The Ti layer sputtered on the silicon nitride
side is used as an adhesive layer for the Au deposition. On the other side of the silicon
nitride substrate, a 10 nm Ir is evaporated (FIG \ref{fig: fab}.a.4). After the evaporation process, a final milling process is done by a FIB. The drilling
process is done by using a $Ga^+$ ions beam and produces the requested nano-scale sinusoidal
pattern. A SEM image of the fabricated grating can be seen in (FIG.\ref{fig: sem_tg}).

\section{Comparison of two STG Design}

Comparison between two theoretical diffraction patterns of the two STGs is presented in linear and logarithmic scales. The numerical calculation has been done in a specific 
wavelength of 13.5 nm.
\begin{figure}
\includegraphics[width=80mm,scale=0.5]{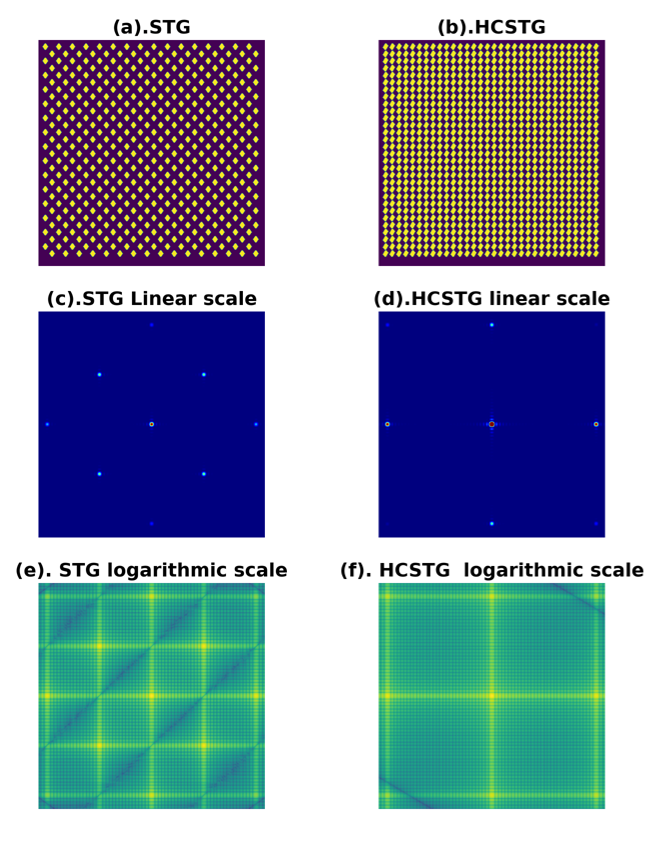}
\caption{\label{fig: comp}Diffraction patterns of HCSTG and STG are 
presented in linear and logarithmic scales. The simulations have been calculated in a
specific wavelength of 13.5 nm.}
\end{figure}

The two STG designs, a regular STG\cite{Shpilman2019}, and our new HCSTG design are shown in FIG. \ref{fig: comp}). The HCSTG presents two significant improvements compared to the old design. Firstly, it achieves a higher transfer efficiency to the first diffraction order. Secondly, the new design successfully eliminates the undesirable "half-order" effects. The "half-order" effect refers to the contribution of diagonal dispersion in the horizontal axis. In the regular STG design, this diagonal contribution lies midway between the zero-order and the first-order, leading to the presence of the "half-order" in the spectrum. However, in the HCSTG design, the diagonal dispersion is precisely aligned with the first order, effectively eradicating the "half-order" contribution.

\subsection{\label{sec:3A}Efficiency}

\begin{figure}
\includegraphics[width=85mm]{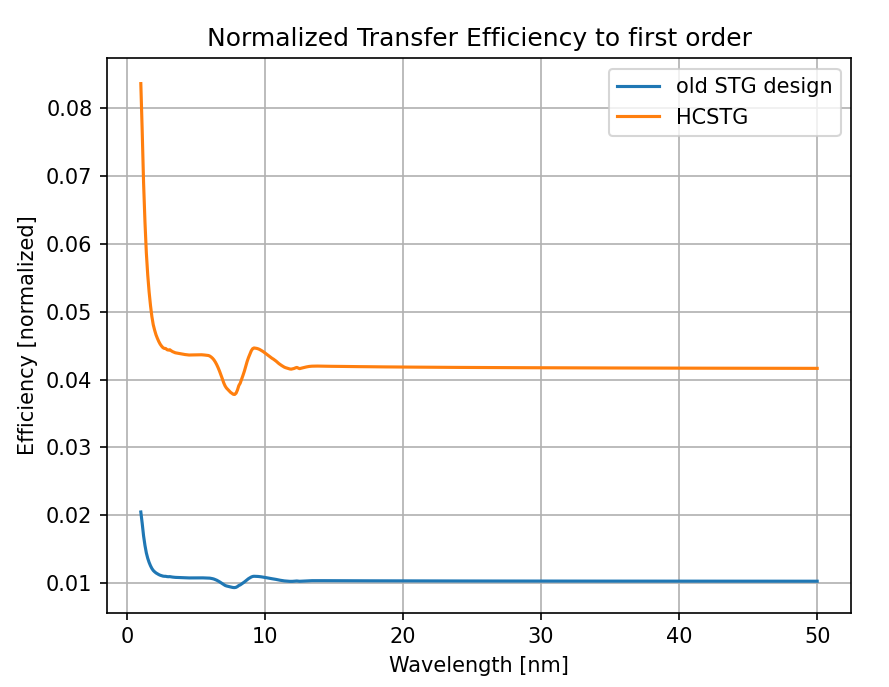}
\caption{\label{fig: eff} Normalized efficiencies of the two STGs designs to first order. The HCSTG design  is more efficient by a factor of 4 compared to the old design}
\end{figure}

High transfer efficiency became possible by reshaping the "eye" shaped apertures. The new hole design allows them to be positioned in a row without losing the stability of the structure because of zeros convergence. In other words, the vertical shifts of the hole tips make it possible to achieve a denser structure with a larger open-to-blocked area ratio without producing a continuous groove that would break the grating. The HCSTG design resulted in a better contrast factor which significantly improves the efficiency to first order of diffraction. As shown in (FIG.\ref{fig: eff}) The HCSTG design has around four times higher transfer efficiency than the previous design to the first diffraction order. This is, of course, a theoretical efficiency that indicates the maximum possible value for an ideally formed STG. Real STG production is a non-trivial procedure, so a drop in the efficiency value shown in (FIG.\ref{fig: eff}) is expected. This manufacturing deviation of an actual grating from the theoretical design is equivalent to both designs, therefore the HCSTG will still be significantly more efficient even after production.

\subsection{\label{sec:3B}Half Diffraction Order Elimination}

The half-orders are caused by the period of the diagonal periodicity of the TG. To better understand this, let's examine the TG's periodicity along three axes: the horizontal axis (our measurement axis), the vertical axis, and the diagonal axis. The diagonal periodicity results in diagonal diffraction, which is responsible for the diagonal first order observed in FIG.\ref{fig: comp}. (c) 
In the old TG design, at a wavelength of 13.5 nm, the diagonal half-order is positioned halfway horizontally between the first horizontal order and the zero order.
Due to the TG's finite shape, all the 
diffraction orders, in each of the three axes, get a "sinc" convolution. This effect means that the diagonal first-order not only "lives" on the diagonal axis, but also contributes to the measurement on the horizontal axis. Consequently, the diagonal first order not only occurs along the diagonal axis but also contributes to the measurement on the horizontal axis, thus leading to the distribution of diagonal orders onto the first order. The distance on the sensor from the zero order to the first order can be calculated from geometric consideration as:

\begin{equation}\label{9}
 x=L_2\cdot m\cdot\frac{\lambda}{d}
\end{equation}

To address this issue, the HCSTG design ensures that the diagonal periodicity is reduced by a factor of $\sqrt{2}$ compared to the horizontal periodicity. Consequently, for each wavelength, the distance $x_d$ of the diagonal first order from the zero order is $\sqrt{2}$ times longer than the distance $x$ of the horizontal first order. This factor of $\sqrt{2}$ in the distances means that, for each wavelength, the diagonal-order distribution on the first order contributes only to that specific wavelength. In addition to the horizontal location of the half-order, it is also further away in the vertical axis. The greater distance significantly reduces the impact of the half-order effect.  

In the logarithmic scale diffraction patterns depicted in FIG.\ref{fig: comp}, a distinct observation arises. In subfigure (f), we notice that the diagonal order aligns horizontally with the first order, which is in contrast to subfigure (e) where the diagonal order lies exactly halfway horizontally, leading to the occurrence of a disruptive half-order phenomenon on the horizontal axis. This alignment disparity stems from the innovative design of the HCSTG, in which the diagonal orders exhibit a period that differs by a factor of $\sqrt{2}$. This unique characteristic of the HCSTG design enables the elimination of the undesirable "half-order" distribution that was prevalent in the previous STG design.   

\section{\label{sec:4}Results}
 
\begin{figure}
\includegraphics[width=80mm,scale=0.5]{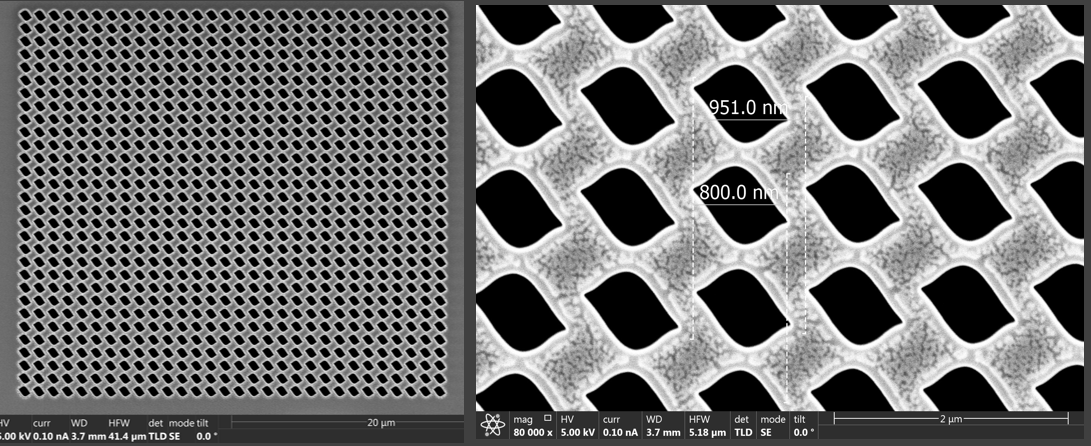}
\caption{\label{fig: sem_tg} SEM images of the HCSTG. A period of 950 nm  }
\end{figure}

\begin{figure}
\includegraphics[width=60mm,scale=0.5]{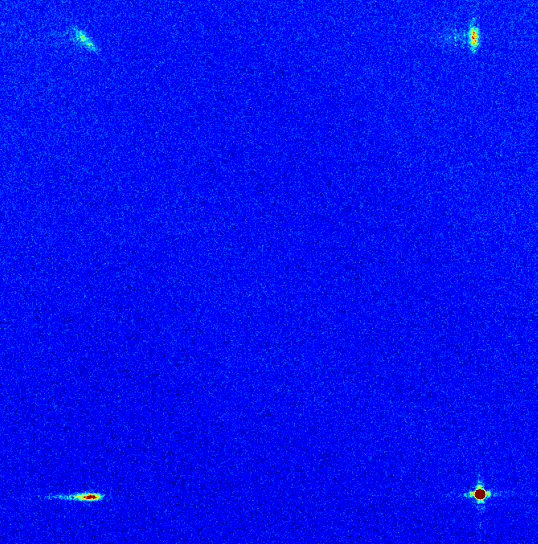}
\caption{\label{fig: cal} The diffraction pattern measured through a multi-layer mirror. The zero order, the first horizontal order, and the first vertical order appear as well as the diagonal order located above the first horizontal order}
\end{figure}

\begin{figure*}
\includegraphics[width=175mm]{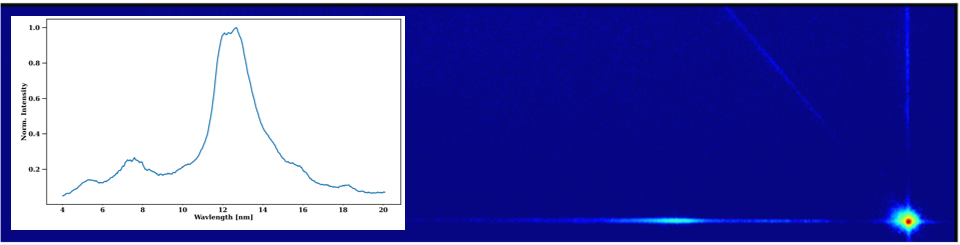}
\caption{\label{fig: spec}The spectrum obtained from the measurement above The diffraction pattern on the CCD sensor. The Zero order and the first order appear as well as part of the diagonal order.  }
\end{figure*}

FIG.\ref{fig: sem_tg} displays a scanning electron microscope (SEM) image depicting the generated HCSTG. This particular HCSTG represents one of the fabricated transmission gratings (TGs) and possesses a period of 950 nm. Similarly, utilizing the methodology elucidated in the fabrication process section above, HCSTGs with periods of 300 nm and 350 nm were also successfully produced. FIG.\ref{fig: cal} portrays the outcomes of the measurement performed on the semi-monochromatic source employed to validate the parameters. In the measurement, as explained in the calibration method section, a narrow spectrum around 13.5 nm was measured. FIG.\ref{fig: cal} presents the diffraction pattern on the CCD sensor. The zero-order and first horizontal, vertical, and diagonal orders appear as expected. Notably, the diagonal order is positioned as intended, precisely above the initial horizontal order.    
The entire obtained spectrum of the laser-produced Sn plasma is shown in FIG.\ref{fig: spec}. The spectrum centered around the wavelength of 13.5 nm is presented on top of the raw data from the spectrometer measurement. 

\section{\label{sec:5}Conclusions}
Previous studies have elucidated the benefits of employing a sinusoidal transmission grating design in contrast to a conventional bar transmission grating design. The HCSTG design that is presented in this article shares all the known advantages of sinusoidal shape transmission grating plus more
useful and novel features such as 4-fold enhancement in efficiency, and half-order elimination. The HCSTG spectrometer presented in this article was specially designed to optimize spectral resolution within the EUV spectral range. Those advantages of HCSTG design enable a very pure and accurate first-order measurement of EUV radiation with no high-order overlapping and no diagonal-order disturbance. In the case of measuring EUV for both academic and industrial research needs, this attribute can be beneficial.

\section{\label{sec:level1}References}
\nocite{*}
\bibliography{library}

\begin{thebibliography}{16}%
\makeatletter
\providecommand \@ifxundefined [1]{%
 \@ifx{#1\undefined}
}%
\providecommand \@ifnum [1]{%
 \ifnum #1\expandafter \@firstoftwo
 \else \expandafter \@secondoftwo
 \fi
}%
\providecommand \@ifx [1]{%
 \ifx #1\expandafter \@firstoftwo
 \else \expandafter \@secondoftwo
 \fi
}%
\providecommand \natexlab [1]{#1}%
\providecommand \enquote  [1]{``#1''}%
\providecommand \bibnamefont  [1]{#1}%
\providecommand \bibfnamefont [1]{#1}%
\providecommand \citenamefont [1]{#1}%
\providecommand \href@noop [0]{\@secondoftwo}%
\providecommand \href [0]{\begingroup \@sanitize@url \@href}%
\providecommand \@href[1]{\@@startlink{#1}\@@href}%
\providecommand \@@href[1]{\endgroup#1\@@endlink}%
\providecommand \@sanitize@url [0]{\catcode `\\12\catcode `\$12\catcode
  `\&12\catcode `\#12\catcode `\^12\catcode `\_12\catcode `\%12\relax}%
\providecommand \@@startlink[1]{}%
\providecommand \@@endlink[0]{}%
\providecommand \url  [0]{\begingroup\@sanitize@url \@url }%
\providecommand \@url [1]{\endgroup\@href {#1}{\urlprefix }}%
\providecommand \urlprefix  [0]{URL }%
\providecommand \Eprint [0]{\href }%
\providecommand \doibase [0]{http://dx.doi.org/}%
\providecommand \selectlanguage [0]{\@gobble}%
\providecommand \bibinfo  [0]{\@secondoftwo}%
\providecommand \bibfield  [0]{\@secondoftwo}%
\providecommand \translation [1]{[#1]}%
\providecommand \BibitemOpen [0]{}%
\providecommand \bibitemStop [0]{}%
\providecommand \bibitemNoStop [0]{.\EOS\space}%
\providecommand \EOS [0]{\spacefactor3000\relax}%
\providecommand \BibitemShut  [1]{\csname bibitem#1\endcsname}%
\let\auto@bib@innerbib\@empty
\bibitem [{\citenamefont {Schupp}\ \emph {et~al.}(2019)\citenamefont {Schupp},
  \citenamefont {Torretti}, \citenamefont {Meijer}, \citenamefont {Bayraktar},
  \citenamefont {Scheers}, \citenamefont {Kurilovich}, \citenamefont {Bayerle},
  \citenamefont {Eikema}, \citenamefont {Witte}, \citenamefont {Ubachs},
  \citenamefont {Hoekstra},\ and\ \citenamefont {Versolato}}]{Schupp2019}%
  \BibitemOpen
  \bibfield  {author} {\bibinfo {author} {\bibfnamefont {R.}~\bibnamefont
  {Schupp}}, \bibinfo {author} {\bibfnamefont {F.}~\bibnamefont {Torretti}},
  \bibinfo {author} {\bibfnamefont {R.~A.}\ \bibnamefont {Meijer}}, \bibinfo
  {author} {\bibfnamefont {M.}~\bibnamefont {Bayraktar}}, \bibinfo {author}
  {\bibfnamefont {J.}~\bibnamefont {Scheers}}, \bibinfo {author} {\bibfnamefont
  {D.}~\bibnamefont {Kurilovich}}, \bibinfo {author} {\bibfnamefont
  {A.}~\bibnamefont {Bayerle}}, \bibinfo {author} {\bibfnamefont {K.~S.}\
  \bibnamefont {Eikema}}, \bibinfo {author} {\bibfnamefont {S.}~\bibnamefont
  {Witte}}, \bibinfo {author} {\bibfnamefont {W.}~\bibnamefont {Ubachs}},
  \bibinfo {author} {\bibfnamefont {R.}~\bibnamefont {Hoekstra}}, \ and\
  \bibinfo {author} {\bibfnamefont {O.~O.}\ \bibnamefont {Versolato}},\
  }\bibfield  {title} {\enquote {\bibinfo {title} {Efficient generation of
  extreme ultraviolet light from nd:yag-driven microdroplet-tin plasma},}\
  }\href {\doibase 10.1103/PhysRevApplied.12.014010} {\bibfield  {journal}
  {\bibinfo  {journal} {Physical Review Applied}\ }\textbf {\bibinfo {volume}
  {12}} (\bibinfo {year} {2019}),\
  10.1103/PhysRevApplied.12.014010}\BibitemShut {NoStop}%
\bibitem [{\citenamefont {Banine}, \citenamefont {Koshelev},\ and\
  \citenamefont {Swinkels}(2011)}]{Banine2011}%
  \BibitemOpen
  \bibfield  {author} {\bibinfo {author} {\bibfnamefont {V.~Y.}\ \bibnamefont
  {Banine}}, \bibinfo {author} {\bibfnamefont {K.~N.}\ \bibnamefont
  {Koshelev}}, \ and\ \bibinfo {author} {\bibfnamefont {G.~H.}\ \bibnamefont
  {Swinkels}},\ }\bibfield  {title} {\enquote {\bibinfo {title} {Physical
  processes in euv sources for microlithography},}\ }\href {\doibase
  10.1088/0022-3727/44/25/253001} {\bibfield  {journal} {\bibinfo  {journal}
  {Journal of Physics D: Applied Physics}\ }\textbf {\bibinfo {volume} {44}}
  (\bibinfo {year} {2011}),\ 10.1088/0022-3727/44/25/253001}\BibitemShut
  {NoStop}%
\bibitem [{\citenamefont {Torretti}\ \emph {et~al.}(2020)\citenamefont
  {Torretti}, \citenamefont {Sheil}, \citenamefont {Schupp}, \citenamefont
  {Basko}, \citenamefont {Bayraktar}, \citenamefont {Meijer}, \citenamefont
  {Witte}, \citenamefont {Ubachs}, \citenamefont {Hoekstra}, \citenamefont
  {Versolato}, \citenamefont {Neukirch},\ and\ \citenamefont
  {Colgan}}]{Torretti2020}%
  \BibitemOpen
  \bibfield  {author} {\bibinfo {author} {\bibfnamefont {F.}~\bibnamefont
  {Torretti}}, \bibinfo {author} {\bibfnamefont {J.}~\bibnamefont {Sheil}},
  \bibinfo {author} {\bibfnamefont {R.}~\bibnamefont {Schupp}}, \bibinfo
  {author} {\bibfnamefont {M.~M.}\ \bibnamefont {Basko}}, \bibinfo {author}
  {\bibfnamefont {M.}~\bibnamefont {Bayraktar}}, \bibinfo {author}
  {\bibfnamefont {R.~A.}\ \bibnamefont {Meijer}}, \bibinfo {author}
  {\bibfnamefont {S.}~\bibnamefont {Witte}}, \bibinfo {author} {\bibfnamefont
  {W.}~\bibnamefont {Ubachs}}, \bibinfo {author} {\bibfnamefont
  {R.}~\bibnamefont {Hoekstra}}, \bibinfo {author} {\bibfnamefont {O.~O.}\
  \bibnamefont {Versolato}}, \bibinfo {author} {\bibfnamefont {A.~J.}\
  \bibnamefont {Neukirch}}, \ and\ \bibinfo {author} {\bibfnamefont
  {J.}~\bibnamefont {Colgan}},\ }\bibfield  {title} {\enquote {\bibinfo {title}
  {Prominent radiative contributions from multiply-excited states in
  laser-produced tin plasma for nanolithography},}\ }\href {\doibase
  10.1038/s41467-020-15678-y} {\bibfield  {journal} {\bibinfo  {journal}
  {Nature Communications}\ }\textbf {\bibinfo {volume} {11}} (\bibinfo {year}
  {2020}),\ 10.1038/s41467-020-15678-y}\BibitemShut {NoStop}%
\bibitem [{\citenamefont {Versolato}(2019)}]{Versolato2019}%
  \BibitemOpen
  \bibfield  {author} {\bibinfo {author} {\bibfnamefont {O.~O.}\ \bibnamefont
  {Versolato}},\ }\bibfield  {title} {\enquote {\bibinfo {title} {Physics of
  laser-driven tin plasma sources of euv radiation for nanolithography},}\
  }\href {\doibase 10.1088/1361-6595/ab3302} {\bibfield  {journal} {\bibinfo
  {journal} {Plasma Sources Science and Technology}\ }\textbf {\bibinfo
  {volume} {28}} (\bibinfo {year} {2019}),\
  10.1088/1361-6595/ab3302}\BibitemShut {NoStop}%
\bibitem [{\citenamefont {Voronov}\ \emph {et~al.}(2011)\citenamefont
  {Voronov}, \citenamefont {Anderson}, \citenamefont {Cambie}, \citenamefont
  {Dhuey}, \citenamefont {Gullikson}, \citenamefont {Salmassi}, \citenamefont
  {Warwick}, \citenamefont {Yashchuk},\ and\ \citenamefont
  {Padmore}}]{Voronov2011}%
  \BibitemOpen
  \bibfield  {author} {\bibinfo {author} {\bibfnamefont {D.~L.}\ \bibnamefont
  {Voronov}}, \bibinfo {author} {\bibfnamefont {E.~H.}\ \bibnamefont
  {Anderson}}, \bibinfo {author} {\bibfnamefont {R.}~\bibnamefont {Cambie}},
  \bibinfo {author} {\bibfnamefont {S.}~\bibnamefont {Dhuey}}, \bibinfo
  {author} {\bibfnamefont {E.~M.}\ \bibnamefont {Gullikson}}, \bibinfo {author}
  {\bibfnamefont {F.}~\bibnamefont {Salmassi}}, \bibinfo {author}
  {\bibfnamefont {T.}~\bibnamefont {Warwick}}, \bibinfo {author} {\bibfnamefont
  {V.~V.}\ \bibnamefont {Yashchuk}}, \ and\ \bibinfo {author} {\bibfnamefont
  {H.~A.}\ \bibnamefont {Padmore}},\ }\bibfield  {title} {\enquote {\bibinfo
  {title} {Fabrication and characterization of ultra-high resolution
  multilayer-coated blazed gratings},}\ }\href {\doibase
  10.1016/j.nima.2010.11.064} {\bibfield  {journal} {\bibinfo  {journal}
  {Nuclear Instruments and Methods in Physics Research, Section A:
  Accelerators, Spectrometers, Detectors and Associated Equipment}\ }\textbf
  {\bibinfo {volume} {649}},\ \bibinfo {pages} {156--159} (\bibinfo {year}
  {2011})}\BibitemShut {NoStop}%
\bibitem [{\citenamefont {Voronov}\ \emph {et~al.}(2012)\citenamefont
  {Voronov}, \citenamefont {Anderson}, \citenamefont {Gullikson}, \citenamefont
  {Salmassi}, \citenamefont {Warwick}, \citenamefont {Yashchuk},\ and\
  \citenamefont {Padmore}}]{Voronov2012}%
  \BibitemOpen
  \bibfield  {author} {\bibinfo {author} {\bibfnamefont {D.~L.}\ \bibnamefont
  {Voronov}}, \bibinfo {author} {\bibfnamefont {E.~H.}\ \bibnamefont
  {Anderson}}, \bibinfo {author} {\bibfnamefont {E.~M.}\ \bibnamefont
  {Gullikson}}, \bibinfo {author} {\bibfnamefont {F.}~\bibnamefont {Salmassi}},
  \bibinfo {author} {\bibfnamefont {T.}~\bibnamefont {Warwick}}, \bibinfo
  {author} {\bibfnamefont {V.~V.}\ \bibnamefont {Yashchuk}}, \ and\ \bibinfo
  {author} {\bibfnamefont {H.~A.}\ \bibnamefont {Padmore}},\ }\bibfield
  {title} {\enquote {\bibinfo {title} {Ultra-high efficiency multilayer blazed
  gratings through deposition kinetic control},}\ }\href
  {http://www.pcgrate.com.} {\bibfield  {journal} {\bibinfo  {journal} {OPTICS
  LETTERS}\ }\textbf {\bibinfo {volume} {37}} (\bibinfo {year}
  {2012})}\BibitemShut {NoStop}%
\bibitem [{\citenamefont {Hambach}, \citenamefont {Schneider},\ and\
  \citenamefont {Gullikson}(2001)}]{Hambach2001}%
  \BibitemOpen
  \bibfield  {author} {\bibinfo {author} {\bibfnamefont {D.}~\bibnamefont
  {Hambach}}, \bibinfo {author} {\bibfnamefont {G.}~\bibnamefont {Schneider}},
  \ and\ \bibinfo {author} {\bibfnamefont {E.~M.}\ \bibnamefont {Gullikson}},\
  }\bibfield  {title} {\enquote {\bibinfo {title} {Efficient high-order
  diffraction of extreme-ultraviolet light and soft x-rays by nanostructured
  volume gratings},}\ }\href@noop {} {\bibfield  {journal} {\bibinfo  {journal}
  {OPTICS LETTERS}\ }\textbf {\bibinfo {volume} {26}} (\bibinfo {year}
  {2001})}\BibitemShut {NoStop}%
\bibitem [{\citenamefont {Golub}(2015)}]{Golub2015}%
  \BibitemOpen
  \bibfield  {author} {\bibinfo {author} {\bibfnamefont {M.~A.}\ \bibnamefont
  {Golub}},\ }\bibfield  {title} {\enquote {\bibinfo {title} {Design of dense
  transmission diffraction gratings for high efficiency},}\ }\href {\doibase
  10.1364/josaa.32.000108} {\bibfield  {journal} {\bibinfo  {journal} {Journal
  of the Optical Society of America A}\ }\textbf {\bibinfo {volume} {32}},\
  \bibinfo {pages} {108} (\bibinfo {year} {2015})}\BibitemShut {NoStop}%
\bibitem [{\citenamefont {Miao}\ \emph {et~al.}(2014)\citenamefont {Miao},
  \citenamefont {Gomella}, \citenamefont {Chedid}, \citenamefont {Chen},\ and\
  \citenamefont {Wen}}]{Miao2014}%
  \BibitemOpen
  \bibfield  {author} {\bibinfo {author} {\bibfnamefont {H.}~\bibnamefont
  {Miao}}, \bibinfo {author} {\bibfnamefont {A.~A.}\ \bibnamefont {Gomella}},
  \bibinfo {author} {\bibfnamefont {N.}~\bibnamefont {Chedid}}, \bibinfo
  {author} {\bibfnamefont {L.}~\bibnamefont {Chen}}, \ and\ \bibinfo {author}
  {\bibfnamefont {H.}~\bibnamefont {Wen}},\ }\bibfield  {title} {\enquote
  {\bibinfo {title} {Fabrication of 200 nm period hard x-ray phase gratings},}\
  }\href {\doibase 10.1021/nl5009713} {\bibfield  {journal} {\bibinfo
  {journal} {Nano Letters}\ }\textbf {\bibinfo {volume} {14}},\ \bibinfo
  {pages} {3453--3458} (\bibinfo {year} {2014})}\BibitemShut {NoStop}%
\bibitem [{\citenamefont {Hurvitz}\ \emph {et~al.}(2012)\citenamefont
  {Hurvitz}, \citenamefont {Ehrlich}, \citenamefont {Strum}, \citenamefont
  {Shpilman}, \citenamefont {Levy},\ and\ \citenamefont
  {Fraenkel}}]{Hurvitz2012}%
  \BibitemOpen
  \bibfield  {author} {\bibinfo {author} {\bibfnamefont {G.}~\bibnamefont
  {Hurvitz}}, \bibinfo {author} {\bibfnamefont {Y.}~\bibnamefont {Ehrlich}},
  \bibinfo {author} {\bibfnamefont {G.}~\bibnamefont {Strum}}, \bibinfo
  {author} {\bibfnamefont {Z.}~\bibnamefont {Shpilman}}, \bibinfo {author}
  {\bibfnamefont {I.}~\bibnamefont {Levy}}, \ and\ \bibinfo {author}
  {\bibfnamefont {M.}~\bibnamefont {Fraenkel}},\ }\bibfield  {title} {\enquote
  {\bibinfo {title} {Advanced experimental applications for x-ray transmission
  gratings spectroscopy using a novel grating fabrication method},}\ }\href
  {\doibase 10.1063/1.4746771} {\bibfield  {journal} {\bibinfo  {journal}
  {Review of Scientific Instruments}\ }\textbf {\bibinfo {volume} {83}}
  (\bibinfo {year} {2012}),\ 10.1063/1.4746771}\BibitemShut {NoStop}%
\bibitem [{\citenamefont {Kuang}\ \emph {et~al.}(2011)\citenamefont {Kuang},
  \citenamefont {Cao}, \citenamefont {Zhu}, \citenamefont {Wu}, \citenamefont
  {Wang}, \citenamefont {Wang}, \citenamefont {Liu}, \citenamefont {Jiang},
  \citenamefont {Yang}, \citenamefont {Ding}, \citenamefont {Xie},\ and\
  \citenamefont {Zheng}}]{Kuang2011}%
  \BibitemOpen
  \bibfield  {author} {\bibinfo {author} {\bibfnamefont {L.}~\bibnamefont
  {Kuang}}, \bibinfo {author} {\bibfnamefont {L.}~\bibnamefont {Cao}}, \bibinfo
  {author} {\bibfnamefont {X.}~\bibnamefont {Zhu}}, \bibinfo {author}
  {\bibfnamefont {S.}~\bibnamefont {Wu}}, \bibinfo {author} {\bibfnamefont
  {Z.}~\bibnamefont {Wang}}, \bibinfo {author} {\bibfnamefont {C.}~\bibnamefont
  {Wang}}, \bibinfo {author} {\bibfnamefont {S.}~\bibnamefont {Liu}}, \bibinfo
  {author} {\bibfnamefont {S.}~\bibnamefont {Jiang}}, \bibinfo {author}
  {\bibfnamefont {J.}~\bibnamefont {Yang}}, \bibinfo {author} {\bibfnamefont
  {Y.}~\bibnamefont {Ding}}, \bibinfo {author} {\bibfnamefont {C.}~\bibnamefont
  {Xie}}, \ and\ \bibinfo {author} {\bibfnamefont {J.}~\bibnamefont {Zheng}},\
  }\bibfield  {title} {\enquote {\bibinfo {title} {An ideal stg},}\ }\href@noop
  {} {\bibfield  {journal} {\bibinfo  {journal} {OPTICS LETTERS}\ }\textbf
  {\bibinfo {volume} {3954}} (\bibinfo {year} {2011})}\BibitemShut {NoStop}%
\bibitem [{\citenamefont {Fan}\ \emph {et~al.}(2015)\citenamefont {Fan},
  \citenamefont {Liu}, \citenamefont {Wang}, \citenamefont {Yang},
  \citenamefont {Wei}, \citenamefont {Zhu}, \citenamefont {Xie}, \citenamefont
  {Zhang}, \citenamefont {Qian}, \citenamefont {Yan}, \citenamefont {Gu},
  \citenamefont {Zhou}, \citenamefont {Jiang},\ and\ \citenamefont
  {Cao}}]{Fan2015}%
  \BibitemOpen
  \bibfield  {author} {\bibinfo {author} {\bibfnamefont {Q.}~\bibnamefont
  {Fan}}, \bibinfo {author} {\bibfnamefont {Y.}~\bibnamefont {Liu}}, \bibinfo
  {author} {\bibfnamefont {C.}~\bibnamefont {Wang}}, \bibinfo {author}
  {\bibfnamefont {Z.}~\bibnamefont {Yang}}, \bibinfo {author} {\bibfnamefont
  {L.}~\bibnamefont {Wei}}, \bibinfo {author} {\bibfnamefont {X.}~\bibnamefont
  {Zhu}}, \bibinfo {author} {\bibfnamefont {C.}~\bibnamefont {Xie}}, \bibinfo
  {author} {\bibfnamefont {Q.}~\bibnamefont {Zhang}}, \bibinfo {author}
  {\bibfnamefont {F.}~\bibnamefont {Qian}}, \bibinfo {author} {\bibfnamefont
  {Z.}~\bibnamefont {Yan}}, \bibinfo {author} {\bibfnamefont {Y.}~\bibnamefont
  {Gu}}, \bibinfo {author} {\bibfnamefont {W.}~\bibnamefont {Zhou}}, \bibinfo
  {author} {\bibfnamefont {G.}~\bibnamefont {Jiang}}, \ and\ \bibinfo {author}
  {\bibfnamefont {L.}~\bibnamefont {Cao}},\ }\bibfield  {title} {\enquote
  {\bibinfo {title} {Single-order diffraction grating designed by trapezoidal
  transmission function},}\ }\href {\doibase 10.1364/ol.40.002657} {\bibfield
  {journal} {\bibinfo  {journal} {Optics Letters}\ }\textbf {\bibinfo {volume}
  {40}},\ \bibinfo {pages} {2657} (\bibinfo {year} {2015})}\BibitemShut
  {NoStop}%
\bibitem [{\citenamefont {Zang}\ \emph {et~al.}(2012)\citenamefont {Zang},
  \citenamefont {Wang}, \citenamefont {Gao}, \citenamefont {Zhou},
  \citenamefont {Kuang}, \citenamefont {Wei}, \citenamefont {Fan},
  \citenamefont {Zhang}, \citenamefont {Zhao}, \citenamefont {Cao},
  \citenamefont {Gu}, \citenamefont {Zhang}, \citenamefont {Jiang},
  \citenamefont {Zhu}, \citenamefont {Xie}, \citenamefont {Zhao},\ and\
  \citenamefont {Cui}}]{Zang2012}%
  \BibitemOpen
  \bibfield  {author} {\bibinfo {author} {\bibfnamefont {H.~P.}\ \bibnamefont
  {Zang}}, \bibinfo {author} {\bibfnamefont {C.~K.}\ \bibnamefont {Wang}},
  \bibinfo {author} {\bibfnamefont {Y.~L.}\ \bibnamefont {Gao}}, \bibinfo
  {author} {\bibfnamefont {W.~M.}\ \bibnamefont {Zhou}}, \bibinfo {author}
  {\bibfnamefont {L.~Y.}\ \bibnamefont {Kuang}}, \bibinfo {author}
  {\bibfnamefont {L.}~\bibnamefont {Wei}}, \bibinfo {author} {\bibfnamefont
  {W.}~\bibnamefont {Fan}}, \bibinfo {author} {\bibfnamefont {W.~H.}\
  \bibnamefont {Zhang}}, \bibinfo {author} {\bibfnamefont {Z.~Q.}\ \bibnamefont
  {Zhao}}, \bibinfo {author} {\bibfnamefont {L.~F.}\ \bibnamefont {Cao}},
  \bibinfo {author} {\bibfnamefont {Y.~Q.}\ \bibnamefont {Gu}}, \bibinfo
  {author} {\bibfnamefont {B.~H.}\ \bibnamefont {Zhang}}, \bibinfo {author}
  {\bibfnamefont {G.}~\bibnamefont {Jiang}}, \bibinfo {author} {\bibfnamefont
  {X.~L.}\ \bibnamefont {Zhu}}, \bibinfo {author} {\bibfnamefont {C.~Q.}\
  \bibnamefont {Xie}}, \bibinfo {author} {\bibfnamefont {Y.~D.}\ \bibnamefont
  {Zhao}}, \ and\ \bibinfo {author} {\bibfnamefont {M.~Q.}\ \bibnamefont
  {Cui}},\ }\bibfield  {title} {\enquote {\bibinfo {title} {Elimination of
  higher-order diffraction using zigzag transmission grating in soft x-ray
  region},}\ }\href {\doibase 10.1063/1.3693395} {\bibfield  {journal}
  {\bibinfo  {journal} {Applied Physics Letters}\ }\textbf {\bibinfo {volume}
  {100}} (\bibinfo {year} {2012}),\ 10.1063/1.3693395}\BibitemShut {NoStop}%
\bibitem [{\citenamefont {Shpilman}\ \emph {et~al.}(2019)\citenamefont
  {Shpilman}, \citenamefont {Hurvitz}, \citenamefont {Danon}, \citenamefont
  {Shussman}, \citenamefont {Ehrlich}, \citenamefont {Maman}, \citenamefont
  {Levy},\ and\ \citenamefont {Fraenkel}}]{Shpilman2019}%
  \BibitemOpen
  \bibfield  {author} {\bibinfo {author} {\bibfnamefont {Z.}~\bibnamefont
  {Shpilman}}, \bibinfo {author} {\bibfnamefont {G.}~\bibnamefont {Hurvitz}},
  \bibinfo {author} {\bibfnamefont {L.}~\bibnamefont {Danon}}, \bibinfo
  {author} {\bibfnamefont {T.}~\bibnamefont {Shussman}}, \bibinfo {author}
  {\bibfnamefont {Y.}~\bibnamefont {Ehrlich}}, \bibinfo {author} {\bibfnamefont
  {S.}~\bibnamefont {Maman}}, \bibinfo {author} {\bibfnamefont
  {I.}~\bibnamefont {Levy}}, \ and\ \bibinfo {author} {\bibfnamefont
  {M.}~\bibnamefont {Fraenkel}},\ }\bibfield  {title} {\enquote {\bibinfo
  {title} {A combined sinusoidal transmission grating spectrometer and x-ray
  diode array diagnostics for time-resolved spectral measurements in laser
  plasma experiments},}\ }\href {\doibase 10.1063/1.5051486} {\bibfield
  {journal} {\bibinfo  {journal} {Review of Scientific Instruments}\ }\textbf
  {\bibinfo {volume} {90}} (\bibinfo {year} {2019}),\
  10.1063/1.5051486}\BibitemShut {NoStop}%
\bibitem [{\citenamefont {Lightman}\ \emph {et~al.}(2019)\citenamefont
  {Lightman}, \citenamefont {Wengrowicz}, \citenamefont {Daniel}, \citenamefont
  {Porat}, \citenamefont {Ehrlich},\ and\ \citenamefont
  {Hurvitz}}]{Lightman2019}%
  \BibitemOpen
  \bibfield  {author} {\bibinfo {author} {\bibfnamefont {S.}~\bibnamefont
  {Lightman}}, \bibinfo {author} {\bibfnamefont {J.~M.}\ \bibnamefont
  {Wengrowicz}}, \bibinfo {author} {\bibfnamefont {E.}~\bibnamefont {Daniel}},
  \bibinfo {author} {\bibfnamefont {E.}~\bibnamefont {Porat}}, \bibinfo
  {author} {\bibfnamefont {Y.}~\bibnamefont {Ehrlich}}, \ and\ \bibinfo
  {author} {\bibfnamefont {G.}~\bibnamefont {Hurvitz}},\ }\bibfield  {title}
  {\enquote {\bibinfo {title} {Comparative analysis for different designs of
  two-dimensional binary sinusoidal transmission gratings},}\ }\href {\doibase
  10.1364/ao.58.007397} {\bibfield  {journal} {\bibinfo  {journal} {Applied
  Optics}\ }\textbf {\bibinfo {volume} {58}},\ \bibinfo {pages} {7397}
  (\bibinfo {year} {2019})}\BibitemShut {NoStop}%
\bibitem [{\citenamefont {Sailaja}\ \emph {et~al.}(1998)\citenamefont
  {Sailaja}, \citenamefont {Arora}, \citenamefont {Kumbhare}, \citenamefont
  {Naik}, \citenamefont {Gupta}, \citenamefont {Fedin}, \citenamefont
  {Rupasov},\ and\ \citenamefont {Shikanov}}]{Sailaja1998}%
  \BibitemOpen
  \bibfield  {author} {\bibinfo {author} {\bibfnamefont {S.}~\bibnamefont
  {Sailaja}}, \bibinfo {author} {\bibfnamefont {V.}~\bibnamefont {Arora}},
  \bibinfo {author} {\bibfnamefont {S.~R.}\ \bibnamefont {Kumbhare}}, \bibinfo
  {author} {\bibfnamefont {P.~A.}\ \bibnamefont {Naik}}, \bibinfo {author}
  {\bibfnamefont {P.~D.}\ \bibnamefont {Gupta}}, \bibinfo {author}
  {\bibfnamefont {D.~A.}\ \bibnamefont {Fedin}}, \bibinfo {author}
  {\bibfnamefont {A.~A.}\ \bibnamefont {Rupasov}}, \ and\ \bibinfo {author}
  {\bibfnamefont {A.~S.}\ \bibnamefont {Shikanov}},\ }\bibfield  {title}
  {\enquote {\bibinfo {title} {A simple xuv transmission grating spectrograph
  with sub- angströmangstr¨angström resolution for laser-plasma interaction
  studies},}\ }\href@noop {} {\bibfield  {journal} {\bibinfo  {journal} {Meas.
  Sci. Technol}\ }\textbf {\bibinfo {volume} {9}},\ \bibinfo {pages}
  {1462--1468} (\bibinfo {year} {1998})}\BibitemShut {NoStop}%
\end{thebibliography}%

\end{document}